\input psfig.tex
%
%
\def\entry#1{
\parshape 3 0pt 16.9 true cm
	    1. true cm 15.9 true cm
	    1. true cm 15.9 true cm #1\hfil}
\tenrm
\magnification=1400
\hoffset= -0.4 true cm
\hsize = 16.9 true cm
\nopagenumbers
\emergencystretch 40pt

\font\twelveb=cmbx12

astro-ph/9609108
\vglue 1 true cm
\centerline{\twelveb RESULTS FROM THE FIRST FLIGHT OF BAM}

\hskip 14 true pt

\centerline{G.~S. Tucker$^1$, H.~P. Gush$^1$, M. Halpern$^1$, W. Towlson$^2$}
\centerline{\it $^1$ Dept. of Physics and Astronomy, University of British
Columbia,}
\centerline{\it Vancouver, BC V6T 1Z1, Canada}
\centerline{\it $^2$ Dept. of Physics and Astronomy, University College
London,}
\centerline{\it London WC1E 6BT, England}

\centerline{
\hbox{\psfig{figure=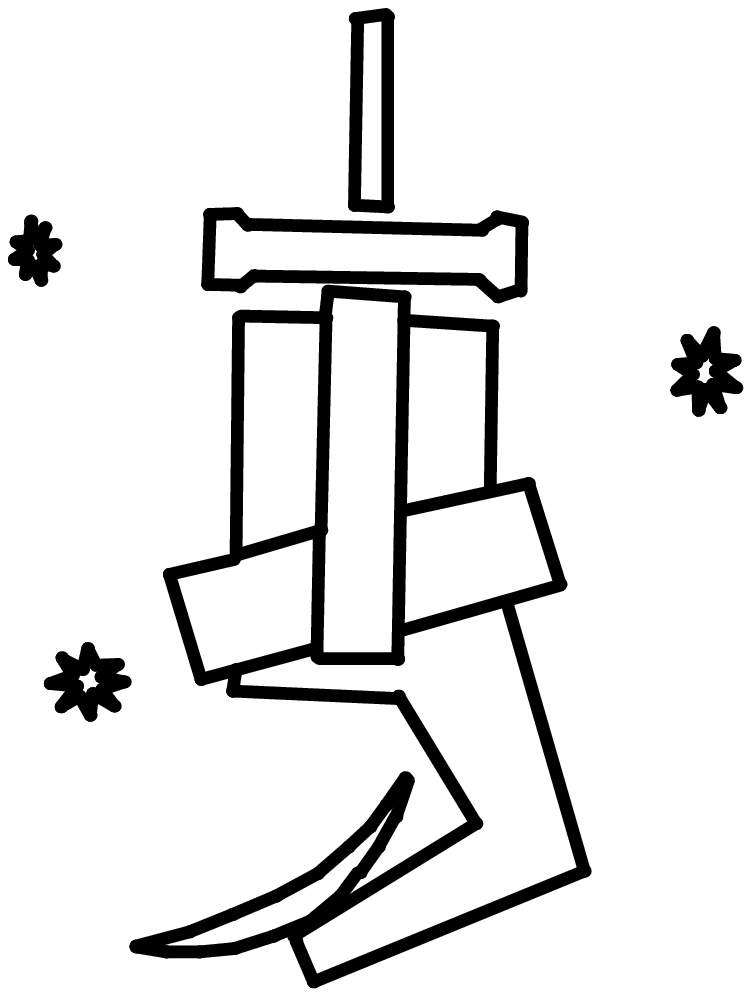,height=8.0in,width=7.2in,rheight=4.8in}}
}

\centerline{\bf Abstract}
\smallskip
\noindent A new instrument, BAM (Balloon-borne Anisotropy Measurement),
designed to measure cosmic microwave background (CMB) anisotropy at
medium angular scales was flown for the first time in July of
1995.  BAM is unique in that it uses a cryogenic differential Fourier
transform spectrometer coupled to a lightweight off-axis telescope.
The very successful first flight of BAM demonstrates the potential of
the instrument for obtaining high quality CMB anisotropy data.
\vfill\eject
\baselineskip=21 true pt

\noindent A new instrument, BAM (Balloon-borne Anisotropy Measurement),
designed to measure the optical spectrum of cosmic microwave
background (CMB) anisotropies was flown for the first time in July
1995.  The instrument is sensitive to anisotropies on angular scales
from 3/4 to a few degrees, scales just larger than those typically
predicted for the first so-called Doppler peak in the anisotropy
angular power spectrum [1].  Therefore, measurements made by BAM will
be used for a sensitive test of the angular power spectrum of {\it
primordial} anisotropies.  In this paper preliminary results obtained
from the first flight are described.

The BAM receiver is a differential Fourier transform spectrometer,
previously used for a measurement of the CMB intensity spectrum from a
sounding rocket [2], [3], coupled to a lightweight prime-focus
telescope.  All of the optical elements of the spectrometer are at
$\sim$2 K except for the bolometric detector assembly, which operates
at 0.26 K.  Rapid scanning of a mirror assembly in the cryostat varies
the optical path length difference in the interferometer, and produces
interferograms at the bolometers whose amplitudes are proportional to
the brightness difference between the two spectrometer inputs.
Spectra are obtained {\it a posteriori} by Fourier transformation of
the interferograms with respect to optical delay.  Useful anisotropy
data are obtained in five spectral channels with central frequencies
from 3.1 cm$^{-1}$ to 9.2 cm$^{-1}$.  Unlike most other instruments
designed to measure the CMB anisotropy, which employ an ambient
temperature chopping mirror to switch the beam on the sky, BAM obtains
a differential measurement with no warm moving optical element; the
only moving optical element is the moving mirror assembly located in
the cryostat.

A diagram of BAM in cartoon form, adapted from the official decal (in
color), is shown on the title page.  The two inputs to the
differential spectrometer are displaced from each other in a direction
perpendicular to the plane of the drawing.  The inputs are located
near to the optic axis of the 70 cm focal length paraboloidal primary
mirror, resulting in two beams $0.7^\circ$ FWHM on the sky separated
by $3.6^\circ$.  Collimators [4] define the acceptance of the two
inputs.  The collimators view the same portion of the primary mirror,
thus the instrument is insensitive to thermal gradients across the
primary mirror.  This was tested late in the flight by heating one
side of the mirror to produce a 2 K gradient across the mirror; no
change in signal is detected.

The BAM gondola structure is relatively large in order to accommodate
the off-axis optical design.  The gondola stands 4 m high in the
laboratory, and the ground shield is 4.5 m across at the widest point.
Nevertheless, the application of aerospace-like design and
construction techniques has produced a lightweight gondola.  For
example, the arm holding the primary mirror is a riveted sheet
aluminum structure similar to an airplane wing.  The undercarriage of
the gondola, which absorbs the impact of landing, is made from
aluminum honeycomb panels supported by a frame made from welded
chrome-molybdenum steel.  The 1.65 m diameter aluminum primary mirror
itself is also lightweight, weighing 26 kg.  The launch weight of BAM
is 660 kg, not including ballast.
  
BAM was launched from the U.S. National Scientific Balloon Facility in
Palestine, Texas at 00:45 UT on 8 July 1995 (sunset on 7 July 1995
local time).  An animation of the launch and other images can be
viewed at http://cmbr.physics.ubc.ca.  The gondola reached a float
altitude of 41.4 km approximately five hours after launch; a little
over three hours was spent performing the experiment at float.  Data
acquisition was terminated when the balloon neared the end of
telemetry range.  The instrument was recovered with only minor damage.

\midinsert
\vbox{
\centerline{
\hbox{\psfig{figure=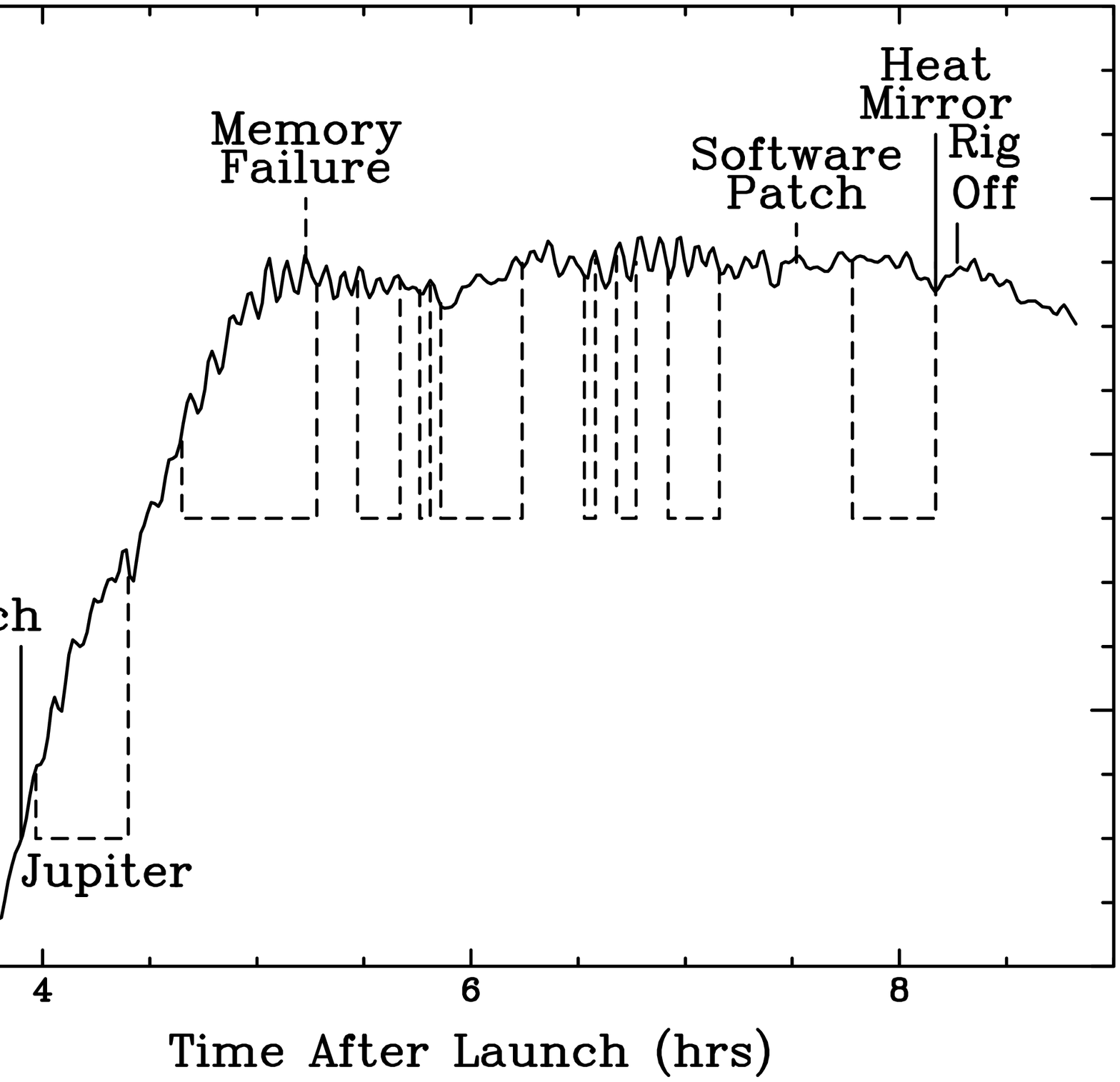,height=2.8in,width=2.8in,rheight=4.0in,angle=-90}
}}
\centerline{\vbox{\baselineskip=14 true pt
\narrower \noindent Figure 1. Chronology of events during the flight.
Jupiter was scanned several times and provides the primary calibration
for the instrument.  Regions indicated by the dashed lines indicate
periods during which data analyzed here were acquired.  The disjoint
nature of these regions is largely due to the problems encountered by
the failure of a memory chip in the pointing system as described in
the text.}}  }
\endinsert

Figure 1 shows a chronology of events during the flight.  Just before
reaching a stable float altitude the telescope was unlatched and
scanned across the planet Jupiter to calibrate the instrument and to
confirm the beam shape.  The instrument was then rotated to look north
and a series of observations was begun near to transit above the north
celestial pole.  Shortly afterwards a memory chip in the pointing
system telemetry electronics became intermittent and then failed,
corrupting some parts of the telemetry from the pointing system.  As a
result, the ground station sent erroneous corrective commands at a
rate exceeding the capabilities of the command transmitter, so
successful command transmission became unreliable.  The telescope
remained locked to guide stars and data were collected, but the
precise sequence of observations was difficult to control.  After
understanding the problem a switch was made to a redundant telemetry
channel not intended for use during flight, and a software patch for
the ground station was developed.  As a result, reliable commanding of
the gondola was restored for the last half hour of observation.

It had been intended to obtain an interleaved set of double difference
measurements by slewing the telescope in azimuth by an angle
corresponding to the beam separation every three minutes.  We were not
entirely successful in this regard.  Since slewing is controlled by
commands from the ground the memory failure limited the number of
double differences obtained.  As a result, preliminary results based
on analysis of single differences are presented; single difference
measurements were obtained on ten fields on the sky.

The spectrum of the optical signal from the sky is contained in the
cosine component of the phase-corrected Fourier transforms of the
interferograms.  The sine component is orthogonal to signals from the
sky and thus provides a monitor of systematic effects in the
measurement.  Using a likelihood analysis technique and assuming that
the sky can be described by a Gaussian autocorrelation function [5],
it is found that the 90\% confidence interval for the cosine component
is not consistent with zero.  Although the peak of the likelihood
function for the sine component is not at zero power, the sine
amplitude is consistent with zero.  The detected optical signal can be
expressed as the square root of the band power or $Q_{\rm flat}$
(after [6]).  It is found that $Q_{\rm flat} = 35.9^{+17.7}_{-6.3}$
$\mu$K at an effective spherical harmonic of $\overline\ell = 74$.
Details can be found in [7].  Figure 2 shows this result along with
the results of other current CMB anisotropy measurements.
     
\midinsert
\vbox{
\centerline{
\hbox{\psfig{figure=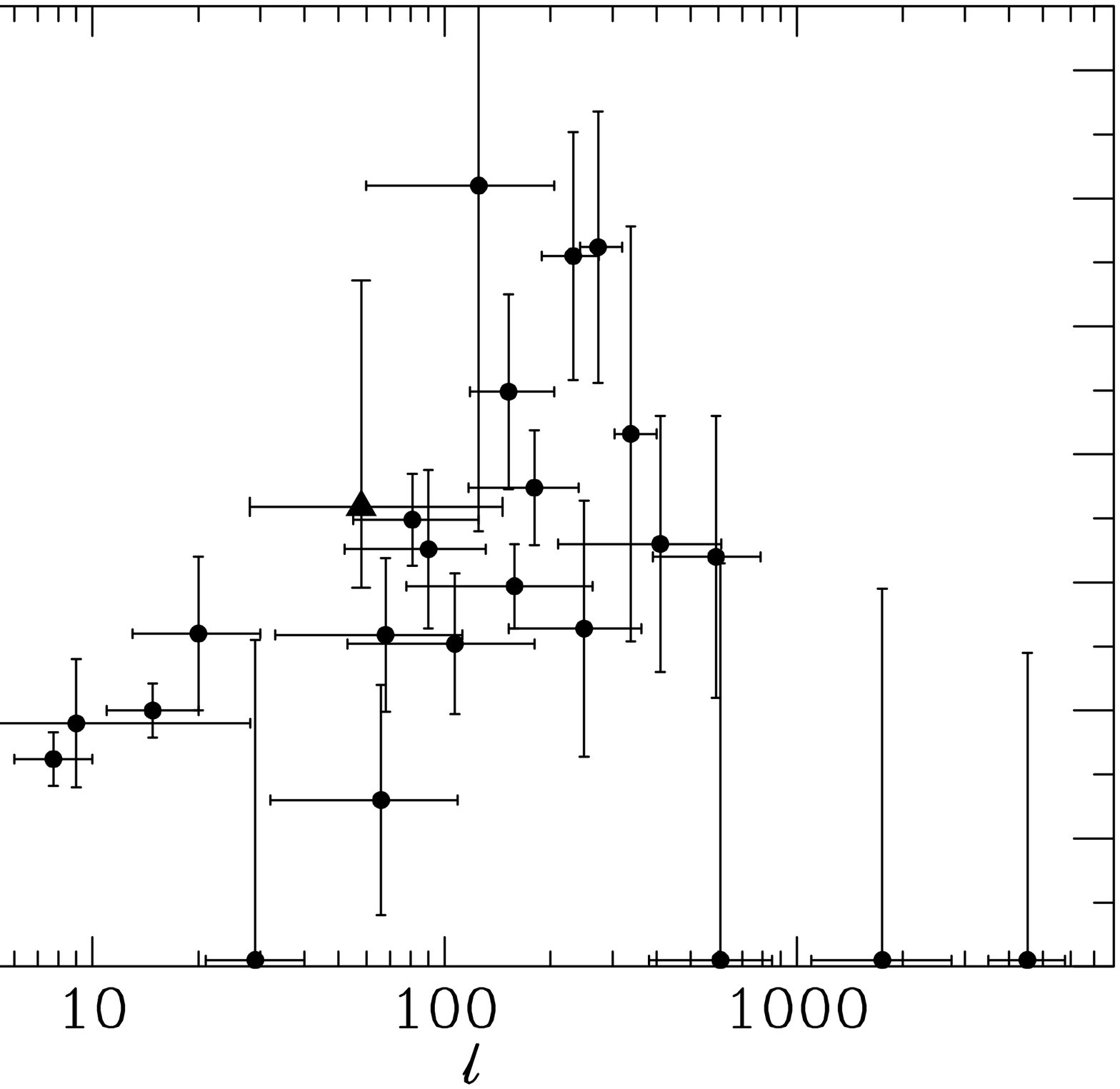,height=2.8in,width=2.8in,rheight=4.0in,angle=-90}
}}
\centerline{\vbox{\baselineskip=14 true pt
\narrower \noindent Figure 2. Current measurements of CMB anisotropy.
The BAM result is shown as a triangle.  Measurements of CMB anisotropy
are a measurement of variance and are thus biased away from zero.
Thus in regions where there are more measurements one expects a larger
scatter as evidenced in this figure.  This figure is adapted from [8].}}  }
\endinsert
   
Statistically significant fluctuations in the microwave sky have been
observed, but not with the sensitivity required to attribute these
fluctuations definitively to cosmic origin; in particular it has not
been possible to constrain the amplitude of the fluctuations and their
optical spectral signature simultaneously.  The modest sensitivity
obtained in this flight is the result of mechanical and electrical
malfunctions, coupled with the short observing time accepted for the
initial flight of a new instrument, and it is not the result of any
fundamental limitation of the instrument.  Avoiding a repeat of these
malfunctions poses no serious technical challenge or risk.  With a
longer flight and some improvements to the receiver, an improvement in
the signal-to-noise ratio of at least a factor of five is
conservatively anticipated.

\medskip

\noindent {\bf Acknowledgements} This research was supported by the Canadian
Space Agency, the Natural Sciences and Engineering Research Council of
Canada, and the Particle Physics and Astronomy Research Council of the
U.K.

\vfill\eject

\noindent {\bf References}
\smallskip
{\frenchspacing
\parfillskip=14 true pt
\parindent = 0pt

\entry{[1] White, M., Scott, D., \& Silk, J. 1994, {\it ARA\&A} {\bf 32}, 329}

\entry{[2] Gush, H.~P., Halpern, M., \& Wishnow, E. 1990, {\it Phys. Rev.
Lett.} {\bf 65}, 537}

\entry{[3] Gush, H.~P., \& Halpern, M. 1992, {\it Rev. Sci. Inst.} {\bf 63},
3249}

\entry{[4] Welford, W.~T., \& Winston, R. 1989, {\it High Collection Nonimaging
Optics}, Academic Press}

\entry{[5] Readhead, A.~C.~S., et al. 1989, {\it Astrophys. J.} {\bf 346}, 566}

\entry{[6] White, M., \& Scott, D. 1994, in {\it CMB Anisotropies Two Years
After COBE}, ed. L. Krauss, World Scientific}

\entry{[7] Tucker, G.~S., Gush, H., Halpern, M., \& Towlson, W. 1996,
{\it submitted to Astrophys. J.}}

\entry{[8] Smoot, G., \& Scott, D. 1996, {\it Phys. Rev. D.}, in press}

}

\bye